# Numerical Simulations of Geomechanical Deformation, Fluid Flow and Reactive Transport in Shale Rough-Walled Microfractures

Morteza Heydari[1], Feng Liang[2], Hui-Hai Liu[2] and Behzad Ghanbarian[3,4,5]

[1] Porous Media Research Lab, Department of Geology, Kansas State University, Manhattan, KS, 66506, United States of America

[2] Aramco Americas: Aramco Research Center–Houston, 16300 Park Row, Houston, TX, 77084, United States of America

[3] Department of Earth and Environmental Sciences, University of Texas at Arlington, Arlington, TX, 76019, United States of America

[4] Department of Civil Engineering, University of Texas at Arlington, Arlington TX 76019, United States of America

[5] Division of Data Science, College of Science, University of Texas at Arlington, Arlington TX 76019, United States of America

**Abstract**

Improving hydrocarbon production with hydraulic fracturing from unconventional reservoirs requires investigating transport phenomena at the single fracture level. In this study, we simulated geomechanical deformation, fluid flow, and reactive transport to understand the effect of hydraulic fracturing treatment on permeability evolution in shale rough-walled fractures. Using concepts of fractional Brownian motion and surface roughness characterizations with laser profilometer, we first generated three rough-walled microfractures consistent with three laboratory experiments (i.e., E4, E5 and E6). After that, the generated microfractures were subjected to a confining pressure in accord with experimental conditions, and geomechanical deformation was simulated. We used the OpenFOAM software package to simulate the fluid flow and permeability. By comparing the simulated permeability values with the experimentally measured ones we found relative errors equal to 28, 15 and 200% respectively for the experiments E4, E5 and E6. After calibration, however, the relative error dropped below 4%. We next simulated the reactive transport using the GeoChemFOAM solver and investigated permeability evolution in the deformed microfractures. We found that after 10 hrs of reactive transport simulations, permeability increased by 47%, on average, in all cases studied here.

## 1. Introduction

Tight and ultra-tight formations, such as tight gas sandstones and shales are distributed around the world with an estimated endowment of about several thousand trillion cubic feet. Such reservoirs have been successfully explored and produced in the United States and, consequently, became a major energy supplier (Ghanbarian et al., 2023). More importantly, advances in horizontal drilling and hydraulic fracturing led to substantial improvement in hydrocarbon production from unconventional reservoirs (Belyadi et al., 2019; Barati and Alhubail, 2020). Despite numerous and recent progress, we are still far from successfully producing oil and gas in shales and tight reservoirs (Han et al., 2024; Chen et al., 2024).

Although it is difficult to precisely estimate the amount of hydrocarbon in place, the value of the recovery factor in tight gas formations is around 25% after a few years of production (Zoback and Kohli, 2019). Several studies reported recovery factors between 2 and 10% in tight oil reservoirs (Hamdi et al., 2018; Wang et al., 2019). Low oil and gas recovery factors in unconventional reservoirs indicate that the major amount of hydrocarbon is not produced. Accordingly, various approaches were proposed to improve production from low- and ultra-low-permeability reservoirs. One promising method is gas Huff and Puff (Sheng, 2015). Carbon dioxide (Ma et al., 2015; Zuloaga et al., 2017) and methane (Sun et al., 2015; Sharma and Sheng, 2018) as well as immiscible and miscible hydrocarbon may be used in the gas Huff and Puff process (Sanchez-Rivera et al., 2015; Ozowe et al., 2020). For recent reviews on the subject, see Zhou et al. (2018) and Iddphonce et al. (2020). Although gas Huff and Puff has been applied in practice by industry, there are still challenges, such as corrosion, asphaltene precipitation, and viscous fingering associated with large-scale applications (Zhou et al., 2018).

Another approach to enhance hydrocarbon production is the use of hydraulic fracturing and application of specific treatments to stimulate tight reservoirs by enlarging apertures of naturally and hydraulically-induced fractures and enhancing well productivity (Holditch, 2006; Liu, 2017; Guo et al., 2021). The low permeability of carbonate-rich tight gas formations necessitates stimulating techniques to improve conductive flow paths in rock matrices (Guo et al., 2017). Acid fracturing is a common technique that increases permeability and transmissivity resulted from the interaction of flow conduits with a reactive fluid (Haghi et al., 2018; Deng and Peters, 2019; Asadollahpour et al., 2019). This technique was found to be effective on shale formations as well when combined with horizontal drilling (Tariq et al., 2020). However, there are certain challenges in improving the permeability of shale reservoirs (Liang et al., 2020). First, it has been found that hydrocarbon production rate decreases rapidly in the first few months (Baihly et al., 2015; Guo et al., 2017), which may be attributed to the closure of microfractures occurring due to increase in effective stress (Matsuki et al., 2001; Liang et al., 2020). Second, it has been observed that around 65 to 90% of fracturing fluid does not flow back to the well, which reduces permeability and production (Soeder, 2017). To overcome these challenges, Liang et al. (2020) proposed using delayed acid generating materials along with micro-proppants in the pad/prepad fluids to minimize the closure of the microfractures in tight carbonate reservoirs and enlarge the fracture aperture. Those authors carried out a series of proof-of-concept laboratory coreflood experiments to test this idea on a single fracture formed by split core, which resulted in recordable permeability enhancement.

Conducting coreflow experiment, however, is a lengthy and very expensive process. Therefore, computational studies were performed to further understand the acid fracturing and investigate its effect on permeability enhancement. Hydraulic fracturing is the process of inducing fractures in rock formations using pressurized fluids. The hydraulic fracturing procedure comprises three key phases: fracture initiation, fracture propagation, and flow-back (Chen et al., 2022; Guglielmi et al., 2023). Acid fracturing is essentially a special case of hydraulic fracturing, primarily employed for carbonate formations (Aljawad et al., 2019). In an acid fracturing experiment, a certain range of processes take place over various temporal and spatial scales (Deng and Peters, 2019), and the overall coupled behavior of which is described by the reactive transport model here.

During hydraulic fracturing, considerable amounts of fracturing fluids are injected into a reservoir, and only a small fraction of the fluids are recovered during the flowback process (Liang et al., 2019). Although the impact of such fluids on geomechanical properties of reservoirs have been addressed (Belyadi et al., 2019), their effects on source rock characteristics, particularly microstructure, porosity, and permeability, require further investigations. In addition, improving and maintaining the connectivity between natural and induced microfractures are crucial to enhancing the well production rate at reservoir scales (Liang et al., 2020).

Surprisingly, there have been relatively few studies that investigated the relationship between chemical alteration of fractures and permeability evolution in unconventional reservoirs (Bandara et al., 2021; Jew et al., 2022). The main objective of this study, therefore, is to provide numerical insights into an improved hydraulic fracturing treatment that enlarges fracture aperture and enhances well productivity in unconventional reservoirs. We aim to: (1) generate rough-walled microfractures based on surface morphology characterization captured by laser profilometer, (2) simulate geomechanical deformation due to confining pressure, (3) simulate fluid flow and liquid permeability in deformed rough-walled microfractures, and (4) model reactive transport numerically to investigate the effect of treatment on permeability evolution. Findings from this study improve our knowledge of reactive transport and permeability evolution at the microfracture level and ability to enhance well production at larger scales.

## 2. Materials and Methods

In this section, we briefly describe three coreflood experiments on shale samples, and then present the numerical methods applied to generate the rough-walled microfractures and simulate geomechanical deformation, fluid flow, and reactive transport in such microfractures.

### 2.1. Experimental methods

**- Materials**

*Rock Samples.* Two types of organic-rich carbonate source rock samples were used in this study. One was an outcrop from the Eagle Ford Shale. Total organic carbon (TOC) for this sample was 5 wt%. The mineral content was determined by X-ray diffraction (XRD), which was mainly composed of calcite (66 wt%), quartz (26 wt%), dolomite (1 wt%), with minor amount of gypsum (2 wt%), pyrite (less than 1 wt%) and clay (4 wt%).

The second rock-sample set was a lime mudstone from a basin in the Middle East. TOC for this sample was about 8 wt%. The mineral content observed with XRD was mainly calcite (92 wt%), dolomite (2 wt%), quartz (3 wt%), pyrite (1 wt%) and less than 2 wt% clay.

*Proppants and Solid Delayed Acid-Generating Materials*. The same materials in Liang et al. (2020) were used in this study: 100-mesh sand (around 150 $\mu$m in diameter) was used as proppant. Polyglycolic acid (PGA) with an average size of 200 $\mu$m was used as a solid delated acid-generating material.

**- Preparation procedure of split core plugs**

Coreflow experiments were conducted using two half-core splits packed with 100 mesh sand and PGA. The half-core splits were prepared by splitting a full core with 1.0 inch in diameter and 1.5 inch in length using a trim saw in the longitudinal direction. Then the half-core split surface was finally trimmed using a target surface trimmer.

**- Surface morphology characterization**

The texture and surface profile of the split-core plug were analyzed using a Nanovea PS50 profilometer. This instrument was designed with leading edge "Chromatic Confocal" optical technology (axial chromatism) and was International Organization for Standardization and American Society for Testing and Materials compliant. The technique measures a physical wavelength directly related to a specific height without using complex algorithms. The surface characterization was conducted for rock samples before and after coreflow experiment to identify the morphology difference caused by the chemical reactions. Surface height was measured with profilometer on six straight lines on the surface, locations of which are demonstrated in Fig. 1. These data were used to calculate surface roughness characteristics including root-mean-square (RMS) height and Hurst exponent, as described in the next section.

**- Packing treatment chemicals on rock surface**

Three experiments with different amounts of chemicals packed as a monolayer between the two half-core plugs were conducted.

*Experiment 4 (E4).* E4 was the same experiment as *Experiment 4* documented in Liang et al. (2020). 83.5 mg of 100-mesh sand and 42.8 mg of PGA were packed as a monolayer packing between the two half-cores of Eagle Ford outcrop shale. The ratio of 100-mesh sand to PGA was about 2:1.

*Experiment 5 (E5).* E5 was packed with 63.5 mg of 100-mesh sand and 32.4 mg of PGA as a monolayer packing between the two half-cores of Middle Eastern shale sample. The ratio of 100-mesh sand to PGA was about 2:1.

*Experiment 6 (E6).* E6 was packed with 93.8 mg of 100-mesh sand and 11.7 mg of PGA as a monolayer packing between the two half-cores of Middle Eastern shale sample. The ratio of 100-mesh sand to PGA was close to 9:1.

**- Experimental measurement of core permeability**

The general procedure was the same as that documented in Liang et al. (2020). The confining pressure was set to 1250 psi for E4 and E5 and 3250 psi for E6. Backpressure was set to 250 psi in all three experiments. To quantify the permeability changes of each testing core assembly (E4, E5, and E6) due to chemical treatment, three core permeabilities for each set were measured. We first estimate the permeability of the split core assembly at room temperature, called the first permeability herein, by flowing through 2% KCl at flow rates of 0.1, 0.15, 0.2 and 0.25 cm$^3$/min, respectively. The permeability value was then calculated using Darcy's law with the pressure drop for the fluid flow and the corresponding flow rates. Next, the split core assembly was packed with intermixed 100-mesh sand and PGA and heated to 250°F. After that, we conducted flow tests with 2% KCl solutions for 40 hours at the rate of 0.2 cm$^3$/min (E4) or 0.1 cm$^3$/min (E5 and E6), respectively. Then the core was cooled to room temperature, and the permeability of the packed split core after treatment (second permeability) was measured by flowing through 2% KCl at flow rates of 1.0, 1.5 and 2.0 cm$^3$/min, respectively. After removing the sand and the undissolved PGA, the permeability of the split core (third permeability) was measured again under room temperature by flowing through 2% KCl at flow rates of 0.1, 0.15, 0.2 and 0.25 cm$^3$/min, respectively. The permeability corresponds to that for the split core assembly with the roughened fracture surface caused by the chemical reactions between PGA and the rock. The measured first and the third permeabilities are the two which will be used as comparison for our numerical simulation. For E4, as documented in Liang et al. (2020), permeabilities changes were from 69 mD to 174 mD due to the surface roughness change, after 40 hours. For E5, permeabilities changes were from 115 mD to 438 mD due to the surface roughness change. For E6, permeabilities changes were from 4.74 mD to 15.2 mD due to the surface roughness change.

## 2.2. Rough-walled microfracture generation

In this study, we assumed that rough surfaces are self-affine and, therefore, can be characterized by two parameters: (1) root-mean-square height ($h_{rms}$) and (2) Hurst exponent ($H$) (Sahimi, 2011). Based on the fractional Brownian motion method, the $h_{rms}$ and $H$ values were determined from the roughness profiles experimentally measured by Liang et al. (2020). For the Hurst exponent, the values were calculated along horizontal ($H_x$) and vertical ($H_y$) directions. To generate microfractures with rough surfaces, we applied concepts of power spectral density (PSD). The power spectrum $C(q)$ of an isotropic surface is defined as (Lang et al., 2016):

$$C(q) = \frac{1}{2\pi^2} \int \langle h(x)h(0)\rangle e^{-iq\cdot x} d^2x \qquad (2)$$

where $q$ is the frequency, $x = (x, y)$, and $z = h(x)$ in which $h$ is the height. Assuming $\langle h \rangle = \bar{h}$, $h(x)$ is shifted with the amount of $\bar{h}$ either downward or upward to obtain $\langle h \rangle = 0$. Eq. (2) may be evaluated using the discrete Fourier transform and radial averaging, as elaborated by Persson et al. (2004). To generate the rough surfaces and rough-wall microfractures consistent with the experiments, we used an open-source code developed by Lang et al. (2016). An ideal spectrum $C(q)$ has the following form:

$$C(q) = C_0 \left(\frac{q}{q_0}\right)^{-2(H+1)} \qquad q_0 \leq q \leq q_1 \qquad (3)$$

where $q_0$ and $q_1$ are respectively the roll-off and cut-off frequencies (Lang et al., 2016). Having the roughness spectrum $C(q)$, the surface may be generated using the following Fourier series (Lang et al., 2016):

$$h(x) = \sum B(q) e^{i(q\cdot x + 2\pi\phi(q))} \qquad (4)$$

where $\phi(q)$ denotes independent uniform random variables. The factor $B(q)$ is given by:

$$B(q) = \frac{2\pi}{L} C(q)^{1/2} \qquad (5)$$

The Hurst exponent, $H$, and root-mean-square roughness height, $h_{rms}$, were used to calculate $C_0$, in Eq. (3), via:

$$C_0 = \frac{Hs}{\pi q_0^2} h_{rms}^2 \qquad (6)$$

in which the parameter $s$ is calculated as $1/s = 1 + H(1 - (q_L/q_0)^2)$.

The dimensions of the generated microfractures were 1.5 in. by 1 in., in accordance with the experiments (Liang et al., 2020). Table 1 lists the Hurst exponent and $h_{rms}$ determined from the original and generated surfaces. For real rough surfaces, it is not possible to have perfect knowledge of the surface topography as a continuous height distribution $h(x, y)$ (Jacobs et al., 2017). Therefore, the average Hurst exponent was determined for three lines along the $x$ direction, and another three lines along the $y$ direction. As observed from Table 1, for experiment 4 (E4), the average values of $H_x$ and $H_y$ are very close, indicating an isotropic rough surface. However, for E5 and E6, $H_x$ and $H_y$ deviate from each other, revealing that the corresponding surfaces were anisotropic in terms of roughness. The workflow for generating the rough surfaces for E5 is demonstrated in Fig. 2. The parameters $q$ and $C$ shown in Fig. 2c represent frequency and power spectrum, respectively.

## 2.3. Numerical simulations

Simulating models needs to include coupled geochemical and geo-mechanical processes to accurately predict the evolution of fracture permeability (Jew et al., 2022). In the following sections, we explain the geomechanical deformation simulations due to confining pressures, fluid flow simulations through rough-walled microfractures, and reactive transport and permeability evolution.

**- Geomechanical deformation**

The contact problem between two rough surfaces may be considered as the equivalent of contact between a rigid realization of a composite surface and an elastic flat body (Lang et al., 2016). The deformable body has a reduced elastic modulus defined as $E^* = E/2(1 - v)$ where $E$ and $v$ are Young's modulus and Poisson's ratio of the original rock joint, respectively. The numerical solution to this nonlinear contact problem is based on the Fast Fourier transformation of the stress-displacement integrals method (Stanley and Kato, 1997), which is implemented in the same open-source package developed by Lang et al. (2016). The deformation between the contacting rough surfaces at the top and bottom of each microfracture was calculated for three pairs of surfaces corresponding to the experiments 4, 5, and 6 (E4, E5, and E6). For all cases, following Zoback and Kohli (2019), we set Young's modulus and Poisson's ratio equal to 40 GPa and 0.2, respectively. In the deformation simulations, the confining pressure was $\sigma = 1250$ psi (8.62 MPa) for E4 and E5 and $\sigma = 3250$ psi (22.41 MPa) for E6.

## - Permeability

To simulate permeability of the generated microfractures, we used the simpleFoam solver of OpenFOAM (Jasak et al., 2007) and numerically solved the three-dimensional Navier-Stokes equations. We assumed that the flow was laminar, incompressible, and steady. An unstructured hexahedral grid was generated for the computational domain using the snappyHexMesh tool in OpenFOAM, as shown in the workflow given in Fig. 2. The refinement level of the cells close to the top and bottom surfaces were chosen such that the grid captured the roughness of the surfaces reasonably. The total number of grid volume elements for the E4, E5, and E6 were around 2.5, 3.5, and 1.7 million, respectively. The boundary conditions for the velocity were uniform at the inlet, zero gradient at the outlet, and no-slip on all other boundaries. For the pressure, we considered atmospheric pressure at the outlet and zero gradient on all other boundaries. Darcy's law was then used to calculate the permeability from the simulated pressure drop:

$$k = \frac{\mu Q}{A(dp/dx)} \qquad (7)$$

To find the pressure gradient, the average pressure was calculated on several slices normal to the axial direction $x$ along the microfracture. The kinematic viscosity in the simulation was $10^{-6}$ m²/s, the cross-sectional area was $A = 507\ mm^2$, in accord with the experiments, and the flow rate was set to $Q = 0.2$ mL/min in the E4, and $Q = 0.1$ mL/min in the E5 and E6.

**- Reactive transport**

To simulate reactive transport in rough-walled microfracture corresponding to the experiments E4, E5 and E6, we used the transportDBSFoam solver, a part of the GeoChemFOAM package, developed by Maes and Menke (2022) and improved Maes et al. (2022) and built around the OpenFOAM. For a review of various reactive transport approaches; see Molins et al. (2021). In a micro-continuum approach, the computational domain is comprised of fluid and solid phases, and the governing equations include flow, transport, and reaction at the fluid-solid interface. The fluid-solid interface is monitored by considering $V_f$ and $V_s$, which represent the volumes of the fluid and solid phases within each control volume $V$. The volume fractions are characterized as ε = $V_f / V$ and $\varepsilon_s = 1 - \varepsilon$. More details about the solver algorithms may be found in Maes et al. (2022).

A structured grid with 4.5 million cells was used for the reactive transport simulations for the same microfracture used for E4, E5 and E6, with the same flow rates. The pH of the acid injected to the fluid was around 4, close to the value used in the experiments conducted by Liang et al. (2020). We should point out that the numerical simulation of reactive transport did not exactly mimic the experiments in which micro-proppants were used. Exact consideration of the effect of micro-proppants was challenging because of the irregular shape of the particles as well as the necessity to develop a solver, which is capable of modeling both reactive flow and a coupled Eulerian-Lagrangian particulate flow model. Hence, for the reactive transport simulations we did not attempt to fully replicate the experiments. Instead, we sought to evaluate a certain method for simulating the reactive flow while neglecting the influence of micro-proppant particles on the flow.

The values of parameters used in the reactive transport simulator transportDBSFoam are listed in Table 3 (Maes et al., 2022). For E4, the calcite density was determined to be 1,789 kg/m³, based on rock samples with 66% calcite. For E5 and E6, the calcite density was calculated as 2,494 kg/m³, as their corresponding rock samples contain 92% calcite. We assumed that the rest i.e., quartz and other minerals did not involve in the reaction.

## 3. Results and Discussion

### 3.1 Geomechanical deformation

For all three cases (E4, E5, and E6), Fig. 3 demonstrates a change in the aperture distribution for the three generated rough-wall microfractures. As can be observed, the surfaces in E6 experienced a much greater deformation compared to those in E4 and E5. Recall that the confining pressure for E6 was nearly 1.6 times greater than that for E4 and E5, which led to much more contact points in the deformed microfracture and consequently a different velocity field.

Results showed that most deformation occurred for the largest apertures (40-60 $\mu m$ in the E4, 60-100 $\mu m$ in the E5, and 30-60 $\mu m$ in the E6). As can be seen in Fig. 3, before deformation the aperture size distributions approximately conform to the normal probability density function. However, after deformation they are right-skewed. Similar numerical (Pyrak-Nolte and Morris, 2000) and experimental (Huo and Benson, 2015) results were reported in the literature. This is because after subjection to some confining pressure, larger apertures become smaller and accordingly the number of larger apertures decreases, while the number of smaller apertures increases.

### 3.2 Permeability simulations

To investigate the effect of grid resolution on simulated permeability, we conducted a grid independence study. Starting with a coarse grid of 440k cells for E4, we refined the grid uniformly in the *x*, *y*, and *z* directions by a factor of 1.5. We repeated the simulations for grids of sizes 700k, 1.06M, 1.6M, 2.4M, and 3.6M, and plotted the simulated permeability versus the grid size in Fig. 4. The grid convergence trend shows that at the size of 2.4M cells, the change in the simulated permeability becomes insignificant (approximately 1%). Therefore, we selected the grid with 2.4M cells as the main grid and used the same setup for grid generation in other cases i.e., E5 and E6.

Dimensionless velocity and pressure contours for two different slices passing through the center of the domain are demonstrated for all cases (E4, E5, and E6) in Fig. 5. $U$ denotes the inlet velocity value, and $u$ represents the velocity in the axial flow direction. The pressure decreases almost linearly in the axial direction ($x$), approaching the reference pressure of zero at the outlet. Preferential flow paths with high velocities, shown with the red color, can be seen in all simulations in Fig. 5. Maximum velocity occurs where the slice height has relatively similar distance from the top and bottom surfaces. However, they formed better in the simulations for E4 and E5 with the confining pressure $\sigma = 1250$ psi than the simulation for E6 with $\sigma = 3250$ psi. Gray regions in the velocity fields shown in Fig. 5 indicate close contact where there was no flow ($u = 0$).

The value of simulated permeability after geomechanical deformation simulations, indicated as $k_{CFD}$, as well as the corresponding error are presented in Table 2 for each case. The errors ranged from 15% for E5 to 200% for E6 (Table 2). As can be seen, the error for E6 is significantly greater than that for E4 and E5. This can be attributed to the fact that due to greater confining pressure in E6 there were substantially more contact points in the microfracture compared to the other two cases. For more accurate and realistic reactive transport simulations, we calibrated the simulated permeabilities, indicated as $k_{ca}$, by tuning the aperture size to achieve a permeability value close to the experimentally measured one. After calibration, the errors ranged from 0.3% for E4 to around 4% for E5. For all three cases, the error after calibration was less than 4%, which shows very good agreement between the calibrated simulations and experiments reported in section 2.1. Table 2 summarizes the experimentally measured permeability ($k_{EXP}$), flow rate ($Q$), inlet velocity ($U$), pressure drop, simulated permeability ($k_{CFD}$), calibrated permeability ($k_{Ca}$), and absolute value of relative error for the simulations for E4, E5, and E6.

### 3.3 Reactive transport simulations

The reactive transport simulations for the three cases E4, E5, and E6 were conducted for only 36000 seconds (or 10 hours) due to substantial computational cost. The change in the permeability is expected to be accompanied by changes in the volume fraction of the fluid relative to total volume, as demonstrated by Ellis and Peters (2016), which may be evaluated by determining the volume fraction field (ε). Changes in the ε field is demonstrated on a slice for three different time steps including 0, 16000 and 36000 seconds for E4, E5, and E6 in Fig. 6, Fig. 7, and Fig. 8, respectively. For all cases, in certain regions, fluid movement was restricted, resulted in minimal to no alteration in the fluid cross-sectional area. In contrast, areas where fluid had flowed freely did experience the effects of the reaction, led to an expansion in their sizes. The evolution of permeability for the three cases is shown in Fig. 9. In this figure we plotted the simulated permeability normalized to its initial value, $k_0$, against the simulation time. It can be seen that in around 10 hours, the normalized permeability $k/k_0$ increased by around 41%, 43%, and 55% for the simulations for E4, E5, and E6, respectively. The initial and final permeability ($k_0$ and $k_{final}$) values are also presented in Fig. 9. We should note that for the reactive transport simulations we used the GeoChemFOAM solver, while for the fluid flow simulations the simpleFoam solver. This is the reason that the $k_0$ values reported in Fig. 9 are slightly different from the $k_{ca}$ values reported in Table 2. In fact, in the reactive transport simulations, as the fluid region is identified by the volume fraction field, and the base mesh is a structured mesh, it has to be very fine to capture the shape of the microfractures accurately. For the reactive transport simulations, we had to use the mesh size of 4.5 million cells, while for the permeability simulations we used 2.4 million cells. Therefore, the captured fluid regions were not exactly the same, which resulted in the differences between $k_0$ in Fig. 9 and $k_{ca}$ reported in Table 2.

As can be seen in Fig. 9, the value of permeability did not greatly change in the first 4000 s. However, beyond that, the permeability started increasing with time nearly exponentially. Similar results were reported by Gharbi (2014) and de Paulo Ferreira et al. (2020). For instance, Gharbi (2014) experimentally investigated the injection of super critical $CO_2$ into Portland limestone samples and reported three regimes of permeability evolution: (1) a plateau in which dissolution did not change connectivity in the pore space, (2) a linear trend in which permeability increased linearly with time, and (3) a dramatic increase by over three orders of magnitude due to substantial increase in connectivity.

Ishibashi et al. (2013) investigated the permeability evolution of fractures in carbonates by including effects of mechanical stress and fluid pH. They conducted experiments using aqueous solutions of ammonium chloride with pH = 5.0, 6.0, 6.1, 6.3, 6.5 and 7 at confining stresses 3, 5 and 10 MPa, respectively. Those authors reported either permeability increase or decrease depending on the combination of confining stress and fluid pH. Generally speaking, they found increasing permeability for pH < 6.1 and decreasing permeability for pH > 6.5 independent of confining stress.

Spokas et al. (2018) addressed the effect of mineralogy on reactive transport and fracture evolution in carbonate-rich caprocks using a two-dimensional fracture model. They numerically simulated mechanical deformation, fluid flow, and acid-driven reactions based on three mineralogy scenarios: a limestone with 100% calcite, a limestone with 68% calcite, and a banded shale with 34% calcite. Their results showed that transmissivity initially increased fastest in rocks with less calcite. Spokas et al. (2018) attributed that to ability to deliver unbuffered-acid downstream faster. They also found that the spatial pattern of less reactive minerals, not abundance, controlled transmissivity and its evolution.

Although for E4, E5 and E6 we found that the permeability increased with the time (Fig. 9), one should bear in mind that chemical interactions between hydraulic fracturing fluids and minerals/organic matter in shales are complex. Such interactions may result in shale minerals dissolution and accordingly porosity and permeability enhancement (Fazeli et al., 2021) or lead to mineral precipitation and consequently porosity and permeability reduction (Herz-Thyhsen et al., 2019; Asgar et al., 2023). Therefore, the competition between these two processes (i.e., mineral dissolution and precipitation) determines whether porosity and permeability increase or decrease (Jew et al., 2022).

The three cases studied here (E4, E5 and E6) differ in initial permeability, flow rate, and flow pathways due to different roughness characteristics. The amount of change in permeability due to the reactive flow is a complex interaction of these factors and we cannot observe any clear relationship here. However, the deviation observed in the trend of E6 from E4 and E5 can be attributed to its much lower permeability and increased number of contact points in comparison to E4 and E5. This distinction sets E6 apart from the latter two, marking a difference in this particular aspect. Although the effect of sand particles was not taken into account in the simulations, the order of magnitude of the permeability change in E4 is consistent with what reported in experiment of Liang et al. (2020) for the same microfracture. Further investigations and numerical simulations are still required to understand permeability evolution in rough-walled microfractures in the presence of sand particles.

## 4. Conclusion

In this study, permeability and its evolution were numerically studied in rough-walled microfractures of shales. Three microfractures were generated using the fractional Brownian motion method based on surface roughness characterization with a laser profilometer. The geomechanical deformation simulations were conducted on the generated rough-walled microfractures. After that, the permeability of the deformed microfractures was numerically simulated using a 3D incompressible steady fluid flow solver and the OpenFoam platform. By comparing the simulated permeabilities to the experimentally measured values, we found relative errors of less than 4%. Reactive transport simulations were conducted using another 3D finite volume-based fluid flow solver, where the three microfractures were influenced by an acidic fluid flow with a pH of 4, and flow rates of 0.1 ml/min and 0.2 ml/min, for 10 hours. As a result of reacting with the fluid flow, the permeability of the microfracture increased by an average of 47%.

**Acknowledgement**

The authors would like to acknowledge Mr. Nam Mai from Aramco Americas for supporting coreflow experiment and surface profilometer measurement.

procedures from fracture injection/shut-in tests. International Journal of Rock Mechanics and Mining Sciences, 170, 105521.

Guo, T., Li, Y., Ding, Y., Qu, Z., Gai, N., & Rui, Z. (2017). Evaluation of acid fracturing treatments in shale formation. Energy & Fuels, 31(10), 10479-10489.

Guo, T., Tang, S., Liu, S., Liu, X., Xu, J., Qi, N., & Rui, Z. (2021). Physical simulation of hydraulic fracturing of large-sized tight sandstone outcrops. SPE Journal, 26(01), 372-393.

Haghi, A. H., Chalaturnyk, R., & Ghobadi, H. (2018). The state of stress in SW Iran and implications for hydraulic fracturing of a naturally fractured carbonate reservoir. International Journal of Rock Mechanics and Mining Sciences, 105, 28-43.

Hamdi, H., Clarkson, C. R., Ghanizadeh, A., Ghaderi, S. M., Vahedian, A., Riazi, N., & Esmail, A. (2018). Huff-n-puff gas injection performance in shale reservoirs: A case study from Duvernay Shale in Alberta, Canada. In Unconventional Resources Technology Conference, Houston, Texas, 23-25 July (pp. 2919–2943).

Han, X., Yu, H., Tang, H., Song, P., Huang, T., Liu, C., & Wang, Y. (2024). Investigation on oil recovery and countercurrent imbibition distance coupling carbonated water with surfactant in shale oil reservoirs. Fuel, 374, 132409.

Herz-Thyhsen, R. J., Kaszuba, J. P., & Dewey, J. C. (2019). Dissolution of minerals and precipitation of an aluminosilicate phase during experimentally simulated hydraulic fracturing of a mudstone and a tight sandstone in the Powder River Basin, WY. Energy & Fuels, 33(5), 3947-3956.

Holditch, S. A. (2006). Tight gas sands. Journal of Petroleum Technology, 58(06), 86-93.
21

Table 1. Average values of $H_x$, $H_y$ and $h_{rms}$ calculated from the measured roughness data as well as determined from the generated rough surfaces for experiments E4, E5 and E6.

| Parameter | E4 | | E5 | | E6 | |
|---|---|---|---|---|---|---|
| | Experiment | Generated | Experiment | Generated | Experiment | Generated |
| Average ($H_x$) | 0.56 | 0.57 | 0.59 | 0.59 | 0.38 | 0.38 |
| Average ($H_y$) | 0.57 | 0.57 | 0.53 | 0.52 | 0.50 | 0.50 |
| $h_{rms}$ ($\mu m$) | 6.16 | 6.10 | 8.89 | 8.90 | 6.39 | 6.40 |

Table 2. Values for the parameters used in the OpenFOAM permeability simulations as well as the experimentally measured permeability values and the corresponding errors for the experiments E4, E5 and E6.

| Case | $k_{EXP}$ (mD) | $Q$ ($m^3/s$) | $U$ (m/s) | $dp/dx$ (Pa/m) | $k_{CFD}$ (mD) | Error (%) | $k_{Ca}$ (mD) | Error (%) |
|---|---|---|---|---|---|---|---|---|
| E4 | 69 | 3.34E-09 | 9.88E-03 | 95743 | 88.1 | 28.1 | 68.8 | 0.3 |
| E5 | 115 | 1.67E-09 | 2.33E-03 | 27577 | 137.3 | 15.0 | 119.4 | 3.9 |
| E6 | 4.74 | 1.67E-09 | 1.21E-02 | 717046 | 13.85 | 200 | 4.6 | 3.1 |

Table 3. The values of various parameters used in the reactive transport simulations.

| Parameter | Value | Unit |
|---|---|---|
| Kinematic viscosity | $1 \times 10^{-6}$ | $m^2/s$ |
| Diffusion coefficient | $5 \times 10^{-9*}$ | $m^2/s$ |
| Inlet flow rate | $0.1\ and\ 0.2$ | ml/min |
| Inlet acid concentration | $0.0126^*$ | kmol/m$^3$ |
| Reaction constant | $8.9 \times 10^{-4*}$ | m/s |
| Stoichiometric coefficient | $2^*$ | (-) |
| Calcite molecular weight | $100^*$ | kg/kmol |
| Calcite density | 1789 (E4)-2494 (E5, E6)** | kg/m$^3$ |
| Kozeny-Carman constant | $10^{-12*}$ | m$^2$ |

*These values are from Table 1 of Maes et al. (2022).

**Updated because 66% (E4) or 92% (E5 and E6) of the solid phase is calcite and the rest of them do not involve in the reaction.

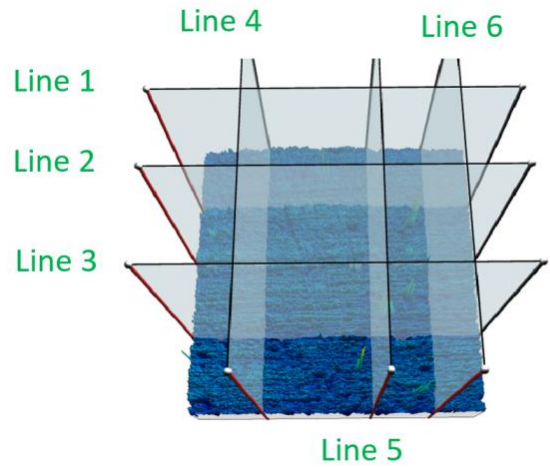

Fig. 1. Three horizonal and three vertical lines used in the experiments to capture the surface roughness by a profilometer for the experiment E4.

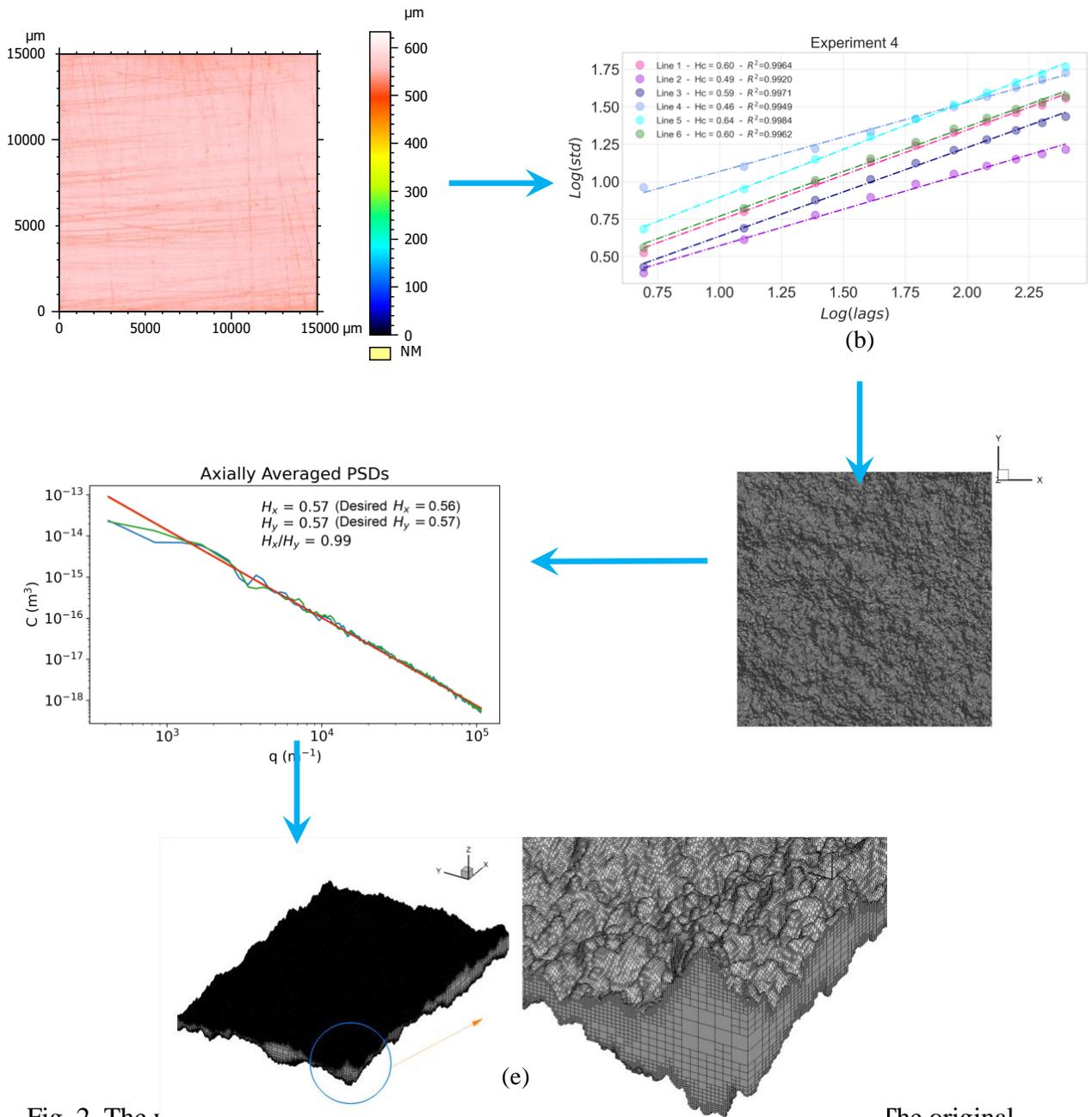

Fig. 2. The workflow for generating rough surfaces for the simulation E4. (a) The original surface, (b) Hurst exponent calculation charts, (c) The generated surface, (d) The desired and resulting Hurst exponent in horizontal and vertical directions, and (e) The CFD mesh generation.

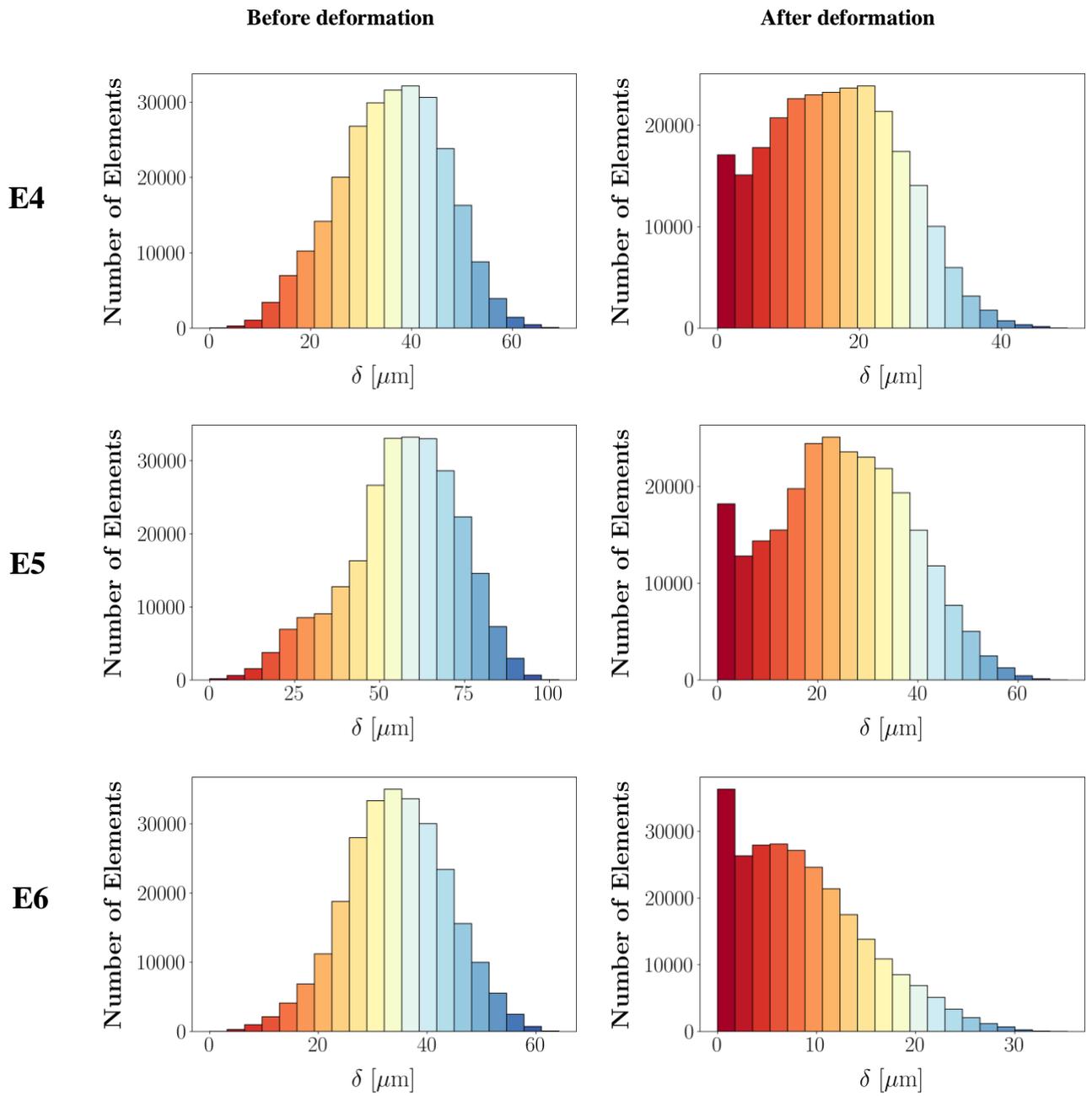

Fig. 3. Aperture size distribution before and after deformation for the simulations E4 (top), E5 (middle), and E6 (bottom).

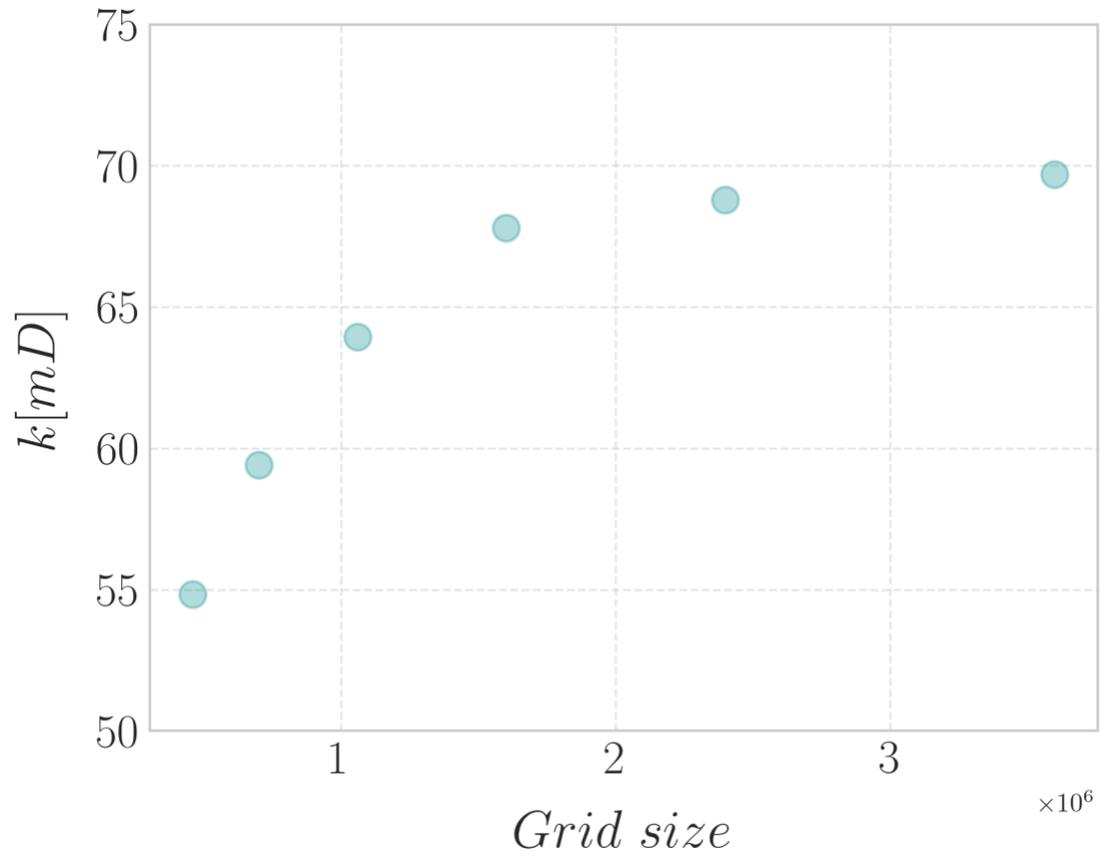

Fig. 4. Simulated permeability against number of cells for the simulation E4. Grid sizes include 440k, 700k, 1.06M. 1.6M, 2.4M, and 3.6M cells.

**E4** 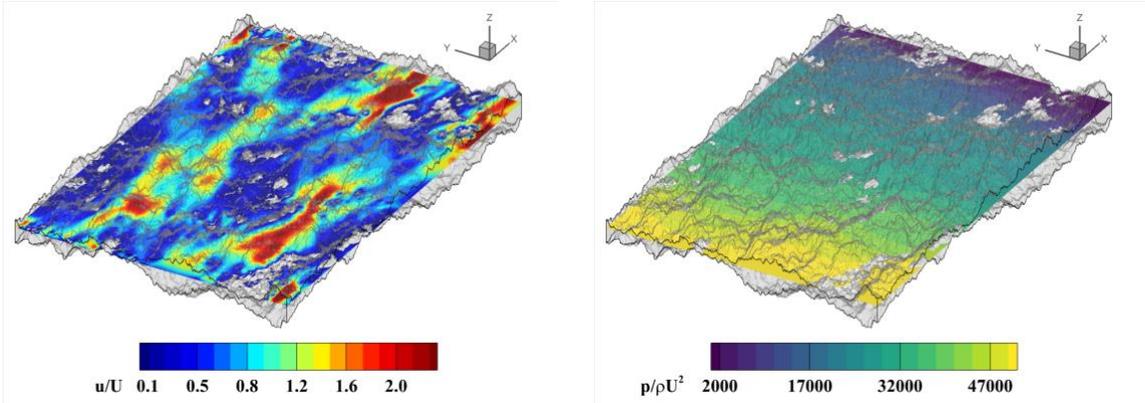

**E5** 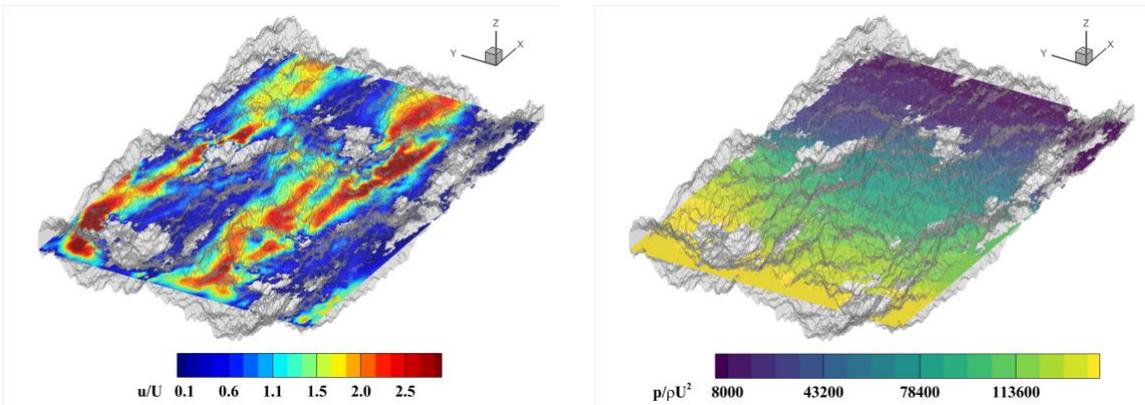

**E6** 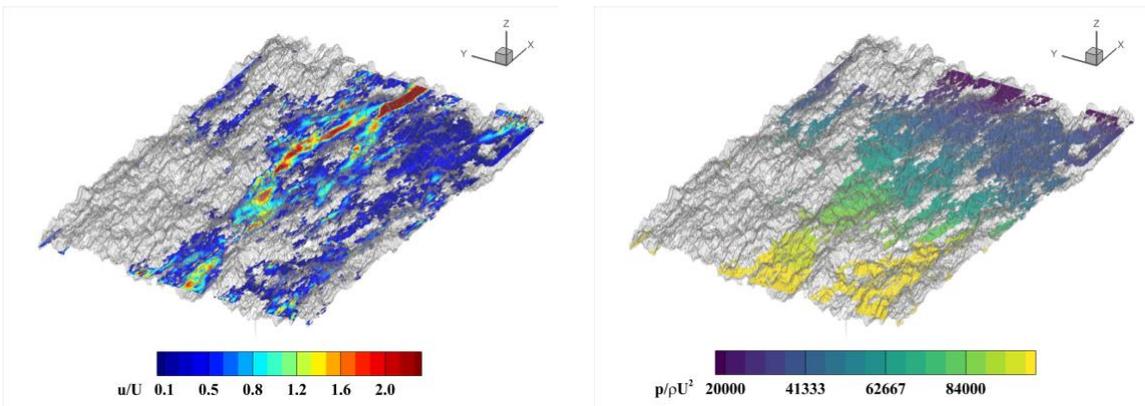

Fig. 5. Dimensionless (left) velocity and (right) pressure contours for one slice perpendicular to the $z$ direction: E4 (top), E5 (middle), and E6 (bottom). $U$ denotes the inlet velocity value, $u$ shows the velocity in the axial flow direction, $p$ is the pressure, and $\rho$ is the fluid density.

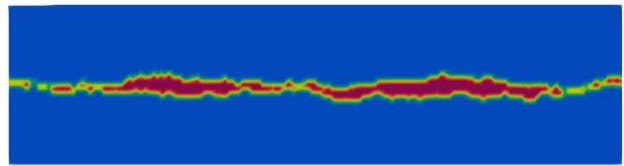

(a)

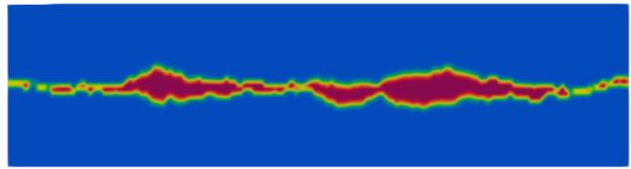

(b)

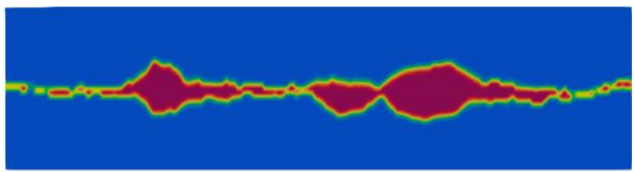

(c)

Fig. 6. The cross section of the microfracture generated for the simulation E4 showing the evolution of the volume fraction field ($\varepsilon$) at the beginning of the simulation time (a) $t = 0$, (b) $t = 16000$ s, and (c) $t = 36000$ s.

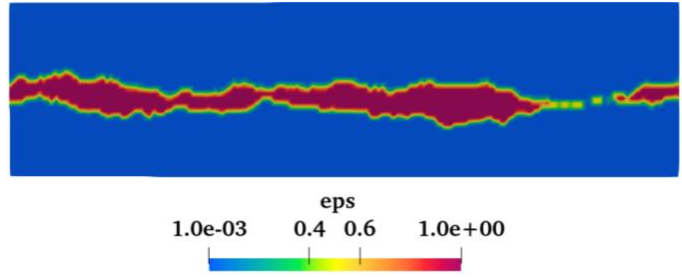

(a)

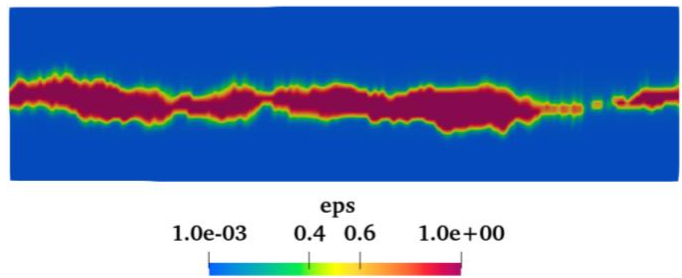

(b)

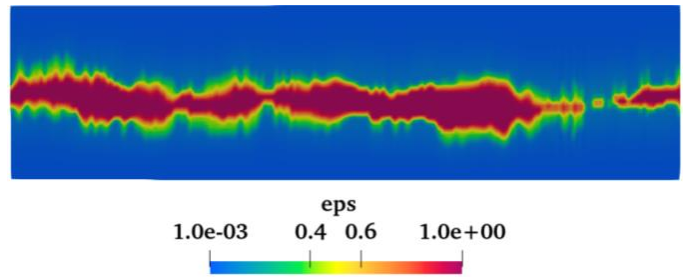

(c)

Fig. 7. The cross section of the microfracture generated for the simulation E5 showing the evolution of the volume fraction field ($\varepsilon$) at the beginning of the simulation time (a) $t = 0$, (b) $t = 16000$ s, and (c) $t = 36000$ s.

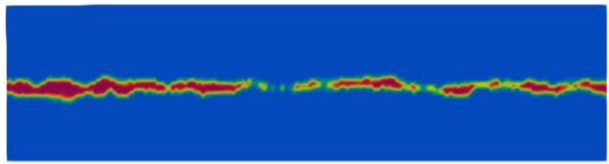

(a)

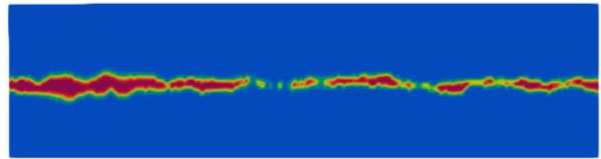

(b)

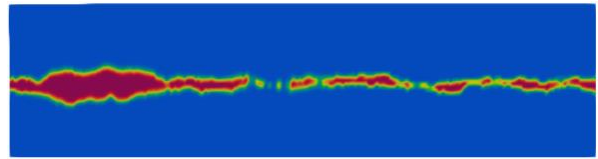

(c)

Fig. 8. The cross section of the microfracture generated for the simulation E6 showing the evolution of the volume fraction field ($\varepsilon$) at the beginning of the simulation time (a) $t = 0$, (b) $t = 16000$ s, and (c) $t = 36000$ s.

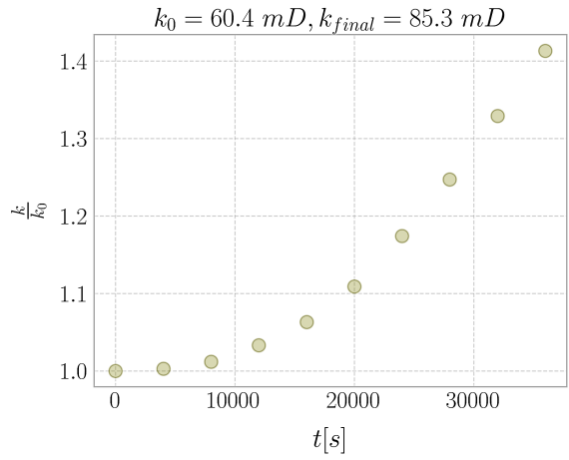
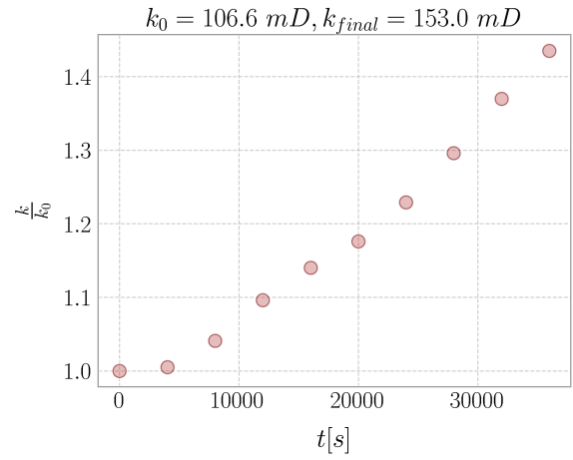

(a)             (b)

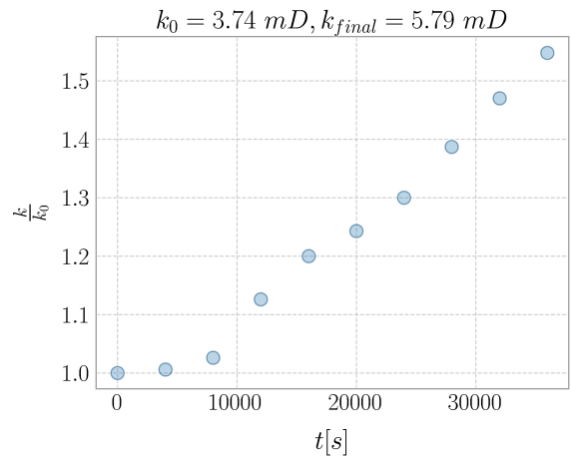

(c)

Fig. 9. Permeability evolution during the reactive transport simulations for the E4 (a), E5 (b), and E6 (c) cases conducted for 10 hours (36000 seconds). Note that the simulated permeability was normalized using its initial value.